\documentclass{article}

\usepackage[preprint]{neurips_2025}

\usepackage[utf8]{inputenc} 
\usepackage[T1]{fontenc}    
\usepackage{hyperref}       
\usepackage{url}            
\usepackage{booktabs}       
\usepackage{amsfonts}       
\usepackage{nicefrac}       
\usepackage{microtype}      
\usepackage{xcolor}         

\usepackage{graphicx}
\usepackage{amsmath,amssymb}
\DeclareMathOperator{\E}{\mathbb{E}}
\usepackage{bm}

\title{A standard transformer and attention with linear biases for molecular conformer generation}

\author{%
  Viatcheslav Gurev \thanks{Corresponding author} \thanks{Equal contribution} \\
  IBM Research \\
  \texttt{vgurev@us.ibm.com} \\
  \And
  Timothy Rumbell \footnotemark[2] \\
  IBM Research\\
  \texttt{thrumbel@us.ibm.com} \\
}

\begin{document}

\maketitle

\begin{abstract}
Sampling low-energy molecular conformations, spatial arrangements of atoms in a molecule, is a critical task for many different calculations performed in the drug discovery and optimization process. Numerous specialized equivariant networks have been designed to generate molecular conformations from 2D molecular graphs. Recently, non-equivariant transformer models have emerged as a viable alternative due to their capability to scale to improve generalization. However, the concern has been that non-equivariant models require a large model size to compensate the lack of equivariant bias. In this paper, we demonstrate that a well-chosen positional encoding effectively addresses these size limitations. A standard transformer model incorporating relative positional encoding for molecular graphs when scaled to 25 million parameters surpasses the current state-of-the-art non-equivariant base model with 64 million parameters on the GEOM-DRUGS benchmark. We implemented relative positional encoding as a negative attention bias that linearly increases with the shortest path distances between graph nodes at varying slopes for different attention heads, similar to ALiBi, a widely adopted relative positional encoding technique in the NLP domain. This architecture has the potential to serve as a foundation for a novel class of generative models for molecular conformations.
\end{abstract}

\section{Introduction}
\label{sec:introduction}

Molecules can adopt a distribution of low-energy states such that the 3D structure of a molecule is best represented by an ensemble of energetically accessible conformations. The conformation distribution informs the possible properties of a molecule, so computing the conformations is of critical importance for drug discovery and optimization applications, including docking, 3D-QSAR, and physicochemical property estimation \cite{hawkins2017conformations}. Therefore, the core problem of molecular conformation generation (MCG) is to sample the conformation distribution to create representative ensembles under limited resource constraints.

MCG is a rapidly developing and promising domain for generative artificial intelligence (AI). Until recently, the primary direction of the research was the development of specialized message passing networks to generate molecular conformations and protein structures. These networks are based on architectures with equivariant bias that use pairwise features, such as distances between atoms, or invariant features derived from local equivariant frames to exploit an invariance of molecular structures to rigid transformations. Notable examples of equivariant models are EGNN  \cite{satorras2021n} and Torsional Diffusion \cite{jing2022torsional}. However, multiple studies demonstrated that non-equivariant models can outperform equivariant networks at MCG. One of the first non-equivariant models, DMCG \cite{zhu2022direct}, led in benchmarks on molecular conformation generation for a relatively long time, with root mean square deviation (RMSD) of $0.69 \text{\AA}$ on the GEOM-DRUGS benchmark, and a recent non-equivariant Molecular Conformer Fields (MCF) model \cite{wang2024swallowing} achieved groundbreaking RMSD of $0.39 \text{\AA}$ on GEOM-DRUGS. A major advantage of non-equivariant models is the ability to utilize domain-agnostic architectures such as the standard transformer or MLP-mixer \cite{tolstikhin2021mlp} models. For instance, MCF processed molecular graphs using the PerceiverIO network \cite{jaegle2021perceiver}, which can handle a variety of input modalities. A domain-agnostic approach is simple and allows reuse of a rich set of architectures, codebases and training infrastructure that have been developed for transformer models.

Despite the success of non-equivariant models, they still exhibit limitations in certain aspects of architecture and performance. MCF obtained SOTA molecule structure generation by using a very large model, by the standards of the field of small molecue structure generation, comprising 242M parameters. Also, while MCF performed well on recall metrics, it showed modest precision results that were subsequently surpassed by a recent flow-matching model, ET-Flow, utilizing an equivariant transformer and harmonic diffusion prior \cite{hassan2024flow}. In this paper, we aim to improve on the performance of non-equivariant models at smaller model sizes to address the shortcoming of requiring large model size. To accomplish this, we designed a new non-equivariant diffusion model using a more advanced positional encoding (PE) compared to SOTA non-equivariant models.

The MCF model used graph Laplacian eigenvectors to represent graph node PE that were fed into PerceiverIO. In standard graph regression benchmarks, a graph transformer with a similar PE \cite{dwivedi2020generalization} has been surpassed by SAN \cite{kreuzer2021rethinking} and Graphormer \cite{ying2021transformers}. SAN uses a learnable PE from eigenvalues and eigenvectors, and Graphormer uses relative PE represented by a learnable attention bias indexed by the shortest path distances. Numerous other options exist for encoding graph structures in transformers; for a review, refer to \cite{rampavsek2022recipe}. It is reasonable to assume that the performance of the non-equivariant model can be improved by using a more advanced PE. Indeed, as we show in this paper, incorporating a simple relative PE with a bias term in self-attention allowed us to create a SOTA model with a small number of parameters that competes with both MCF and ET-Flow models.

We further demonstrate that the difference seen in precision metrics between MCF and ET-Flow models is partially due to different treatment of stereochemistry information, specifically molecular chirality. Chirality tags were directly used as input features when training ET-Flow, but were absent in the MCF model. ET-Flow also used an additional chirality correction step to further improve metric performance. Here, we demonstrate that adding a chirality correction step to a non-equivariant model reduces the deficit of non-equivariant models in precision performance.

To train the model on a limited computational budget, we developed a two-stage training process. At the first stage, the model is trained on molecules with removed hydrogen atoms, which represent on average $\approx 44\%$ of the atoms in molecules in the GEOM-DRUGS dataset. At the second stage the model is finetuned on non-modified molecules (i.e. with hydrogens re-introduced). This training protocol accelerates model training by reducing the number of tokens used to represent molecules in the first stage of training, and its successful application unlocks opportunities for constructing new cascade diffusion models \cite{ho2022cascaded} for molecular conformations. In addition, we explored other aspects of non-equivariant diffusion models that have not attracted attention in the literature before. Most non-equivariant models employ raw atom coordinates in their feature set. As shown in other domains \cite{gorishniy2022embeddings}, small MLP and sinusoidal continuous value encoders can enhance model convergence and improve model performance. Therefore, we investigated if continuous value encoders could additionally improve the quality of generated molecular conformations.

Our key contributions can be summarized as follows
\begin{enumerate}
    \item We developed a new transformer model for MCG with new relative positional encoding that demonstrates state-of-the art recall results for small and medium size models on the GEOM-DRUGS benchmark.
    \item We highlight the importance of including/excluding chirality information to make fair comparisons between models and demonstrate that chirality correction for non-equivariant transformers can improve precision metrics.
    \item Our simple two stage training protocol (removing/restoring hydrogen atoms) makes more efficient use of a limited computational budget when scaling models.
\end{enumerate}

\section{Related Work}
\label{sec:relatedwork}

In this section we discuss the alternative methods to AI for MCG, briefly review ML models for MCG, and discuss the use of positional encoding in graph transformers. Finally, we provide a detailed overview of GEOM dataset benchmarks and how they have been used in MCG models from the literature.

\paragraph{MCG methods.} Conformer ensemble generation methods are usually classified as either stochastic or systematic, based on the type of search used \cite{hawkins2017conformations}. The most complex stochastic methods are molecular dynamics (MD) simulations, which simulate atoms using force-fields and are computationally expensive \cite{riniker2015better}. Enhanced sampling methods, such as metadynamics, can help MD simulations explore more efficiently. The CREST software \cite{pracht2020crest1,pracht2024crest2} provides a suite that includes MD and metadynamics, and has been used to generate reference conformer ensembles in small molecule datasets such as GEOM \cite{axelrod2022geom} that are used as targets to benchmark conformer generation approaches. Alternative stochastic methods can improve speed by searching a lower-dimensional space than MD, using Monte Carlo search or genetic algorithms (GA). These methods can be subject to bias introduced by seed coordinates \cite{hawkins2017conformations}. The distance geometry (DG) approach \cite{havel1983dg} is used by many stochastic algorithms and helps to avoid seed bias by initializing coordinates based on distance constraints, and guides search towards conformations that match these predicted distances. 
Systematic, or rule-based, approaches typically divide the molecule into fragments and apply rules for how each section should conform, using tools like fragment databases or torsion dictionaries, then reassemble the molecule \cite{hawkins2017conformations}. Popular rule-based approaches include OMEGA \cite{hawkins2010omega} and Frog2 \cite{miteva2010frog2}. While purely rule-based approaches are fast, reliance on fragment templates creates difficulties for large or flexible molecules \cite{ganea2021geomol}. Successful methods often combine multiple types of algorithm, for example, Balloon \cite{vainio2007balloon} initializes GAs using DG, and ETKDG \cite{riniker2015better} augments the popular cheminformatics software package RDKit implementation of DG generation (ETDG) with fragment knowledge. A new avenue for MCG came from deep learning, with early models such as G-SchNet \cite{gebauer2019gschnet} based on architectures for molecule property prediction. Recently diffusion models \cite{ho2020denoising} were adapted to generate distributions of conformers \cite{xu2022geodiff}. Subsequently, diffusion models were able to outperform reference stochastic (ETKDG) and rule-based (OMEGA) methods at recreating CREST ensembles \cite{jing2022torsional}.

AI models for conformer generation can be categorized into models for direct MCG and models that produce spatial features of the molecular graph such as inter-atomic distances, which are subsequently used to reconstruct atom coordinates with Euclidean distance geometry algorithms \cite{simm2020generative}. 
Numerous other graph models were developed for predictive and generative molecular analysis; for a recent review, refer to \cite{duval2023hitchhiker}. The majority of these models operate in the Euclidean space of atom coordinates. An exception is Torsional Diffusion \cite{jing2022torsional}, which operates on spaces of torsional angles that have a much lower dimensionality. High numbers of iterations during inference became a restrictive factor for diffusion models. One of the solutions to the problem is harmonic diffusion \cite{jing2023eigenfold}, which initially generates distributions of molecule-like objects with bond lengths close to real molecules and can then be used as a prior for diffusion models. ET-Flow \cite{hassan2024flow}, a flow-matching model trained on such a harmonic prior, requires only 50 iterations over the network during the inference stage.

The first model for conformer generation without equivariant bias, CVGAE \cite{mansimov2019molecular}, was a variational autoencoder with message passing networks. 
Specific to 3D molecular modeling was a rigid alignment of molecules between input and output in the autoencoder loss. 
DMCG \cite{zhu2022direct} improved on the CVGAE architecture, iteratively updating conformations from layer to layer in the network, using advanced modules such as graph attention \cite{brody2021attentive} and considering permutation invariance of symmetric atoms in addition to the alignment in the autoencoder loss. 

A significant paradigm shift occurred due to a recent publication. Instead of specialized graph neural networks, MCF \cite{wang2024swallowing} employed a universal transformer Perceiver IO \cite{jaegle2021perceiver} designed with predictive capabilities for multiple modalities. Utilizing standard architectures helps avoid model tuning and extensive hyperparameter searches while enhancing performance and scalability. The experiments in the MCF paper highlighted an important finding: sample quality improves with increasing model size. This opens possibilities to enhance MCG with standard AI techniques, first by scaling a model to large sizes, then reducing the model size through methods such as quantization and distillation.

\paragraph{Positional encoding in graph transformers.}The MCF model uses global PE, where each atom is assigned a vector comprising coordinates of the first $k$ non-trivial eigenvectors in the eigenbasis of the molecular graph's normalized Laplacian, ordered by ascending eigenvalues. This choice is likely dictated by the MCF problem statement to formulate the model as a diffusion probabilistic field \cite{zhuang2023diffusion}. However, more advanced PE schemes have been developed in the literature \cite{rampavsek2022recipe}. We considered and tested multiple options, including transformers combined with GNNs \cite{min2022transformer}, ultimately selecting a relative PE approach.

Popular relative positional encoding in language models are RoPE \cite{su2024roformer} and ALiBi \cite{press2021train}. In graphs, relative positional encoding can be represented by a bias term in attention that is calculated from distance between nodes. In Graphormer \cite{ying2021transformers}, learnable biases $b_{\phi(v_i, v_j)}$ are employed, indexed by shortest path distance $\phi(v_i, v_j)$ between graph nodes $v_i$ and $v_j$, 
\begin{equation} \label{eq:graphormer}
    A_{ij} = \frac{(\mathbf{h}_i\mathbf{W}_Q)(\mathbf{h}_j\mathbf{W}_K)^T}{\sqrt{d}}+b_{\phi(v_i, v_j)},
\end{equation}
where $A_{ij}$ is $(i, j)$ - element of the Query-Key product matrix $\mathbf{A}$, and $\mathbf{h}_i \in \mathbb{R}^{1\times d}$ is a transformer hidden state of dimension $d$. The mechanism of learnable bias is attractive, as it allows bias variation across different transformer heads. For MCG, heads with a large bias can focus on encoding the graph, while heads with a minimal bias can primarily process atom coordinates.

\paragraph{Datasets and benchmarks for conformation prediction.} The GEOM datasets \cite{axelrod2022geom} have been adopted for evaluation and comparison of machine learning methods for conformer generation. GEOM-DRUGS consists of 304,466 mid-sized molecules (averaging 44.4 atoms per molecule), and the GEOM-QM9 consists of 133,258 small molecules (averaging 18.0 atoms per molecule). Each data entry consists of a SMILES string and an ensemble of 3D conformers generated by simulation using the CREST method \cite{pracht2020crest1, pracht2024crest2}. When training on these datasets, the goal is to train a model that reproduces the distribution of 3D conformations for each SMILES (or 2D graph). To evaluate model-generated conformations and compare distributions, RMSD is calculated between all reference and generated conformations for a molecule, from which two metrics are computed, coverage (COV) and average minimum RMSD (AMR) \cite{zhou2023deep}, according to
\begin{equation} \label{eq:RMSD}
    \mathrm{RMSD}(R, \bar{R}) = \min_{\Phi}\left(\frac{1}{n}\sum_{i=1}^n || \Phi(R_i) - \bar{R_i}||^2\right)
\end{equation}
\begin{equation} \label{eq:COV}
    \mathrm{COV}(S_g, S_r) = \frac{|\{R\in S_r|\mathrm{RMSD}(R, \bar{R})<\delta, \bar{R}\in S_g\}|}{|S_r|}
\end{equation}
\begin{equation} \label{eq:AMR}
    \mathrm{AMR}(S_g, S_r) = \frac{1}{|S_r|}\sum_{R\in S_r} \min_{\bar{R} \in S_g} \mathrm{RMSD}(R, \bar{R}),
\end{equation}
where $S_g$ and $S_r$ are the set of generated and ground-truth conformations for a molecule, respectively; $R$ and $\bar{R}$ are matrices of reference and generated conformations with the heavy atom coordinates in matrix rows, and $\Phi$ are the rigid transformations to align the generated conformation with the reference. In addition to the above recall metrics, precision metrics are usually reported, wherein $S_g$ and $S_r$ switch places in \eqref{eq:COV}-\eqref{eq:AMR}. In all preceding studies, these metrics are calculated on $2k$ generated conformers for each molecule, where $k$ is the number of conformers for a molecule in the test set.

Early SOTA on reproducing GEOM conformations was set by CGCF \cite{xu2021learning}, which trained on only 50,000 conformations from each dataset and evaluated against conformations from 100 molecules. This work established the coverage threshold $\delta$ in \eqref{eq:COV} as 0.5\r{A} and 1.25\r{A} for GEOM-QM9 and GEOM-DRUGS, respectively. Subsequently, ConfGF \cite {shi2021learning} established SOTA after training with 5 conformations from each of 40,000 molecules from both GEOM datasets (200,000 conformations in training sets), and a test set of 200 molecules with between 50 and 500 conformations (22,408 total conformations for GEOM-QM9 and 14,324 for GEOM-DRUGS). Both GeoDiff \cite{xu2022geodiff} and DMCG \cite{zhu2022direct} used these training and test sets, which the latter termed ``small-scale GEOM-DRUGS'' and also trained on a larger set of 2,000,000 conformations from GEOM-DRUGS, furthering the benchmark SOTA. An earlier study that considered every molecule in the GEOM-DRUGS dataset was GeoMol \cite{ganea2021geomol}, which established dataset indices for train/validation/test splits that have been reused in subsequent studies. The splits used 80\%/10\%/10\% of the data, with the test split further downsampled to 1000 random molecules, creating splits containing 106,586/13,323/1,000 and 243,473/30,433/1,000 molecules for train/validation/test for GEOM-QM9 and GEOM-DRUGS, respectively. During training, a random sample of 10 and 20 conformations per molecules were used, for GEOM-QM9 and GEOM-DRUGS, respectively, while every conformer is used from the test split molecules during evaluation.

With Torsional Diffusion \cite{jing2022torsional}, a quantitative leap in the GEOM-DRUGS benchmark was achieved, such that the coverage threshold $\delta$ in \eqref{eq:COV} had to be reduced from 1.25\r{A} to 0.75\r{A} to observe differences between models. This and subsequent studies MCF \cite{wang2024swallowing}, which established current SOTA for recall, and ET-Flow \cite{hassan2024flow}, which established current SOTA for precision, all used the GeoMol data splits. However, comparisons between methods are still confounded as Torsional Diffusion and ET-Flow trained with 30 conformers per molecule for GEOM-DRUGS and GEOM-QM9, while MCF used 20 and 10 conformers for GEOM-DRUGS and GEOM-QM9, respectively, established by GeoMol.

The process for generating reference conformers in the GEOM datasets involves a graph re-identification step \cite{axelrod2022geom} that can assign different graphs to the same molecule, which are grouped according to the original SMILES. Therefore, RMSDs cannot necessarily be calculated between all reference and generated conformations for a molecule in the dataset. Torsional diffusion \cite{jing2022torsional} simply dropped any such molecules from data splits. MCF \cite{wang2024swallowing} generated $2$ conformations for each reference conformation, but, because RMSD can only be computed for conformers with identical graphs, some molecules effectively have a reduced number of RMSD values, altering the metrics and confounding comparisons unless identical methods for handling these discrepancies are used. Here we followed the MCF approach, grouping conformers by reference molecule.

\paragraph{Molecule chirality.}
In the GEOM dataset all conformers of a given molecule have the same chirality. Thus, for optimal performance it may be important to include chirality in the feature set of 2D molecular graphs, as was done in Torsional Diffusion and ET-FLow. Torsional Diffusion generated conformations with the correct chirality by design. ET-Flow used chirality tags in the atom feature set and chirality correction step. However, we found that ET-Flow \textit{SO(3)} (without explicit chirality correction step) generates ensembles of conformers for a given molecule with a distribution of chiralities. Similar observation was made for MCF model. We demonstrate here that chirality correction significantly improve model performance and should be accounted in fair model comparison.

\begin{figure}
  \centering
  \includegraphics[width=.79\linewidth]{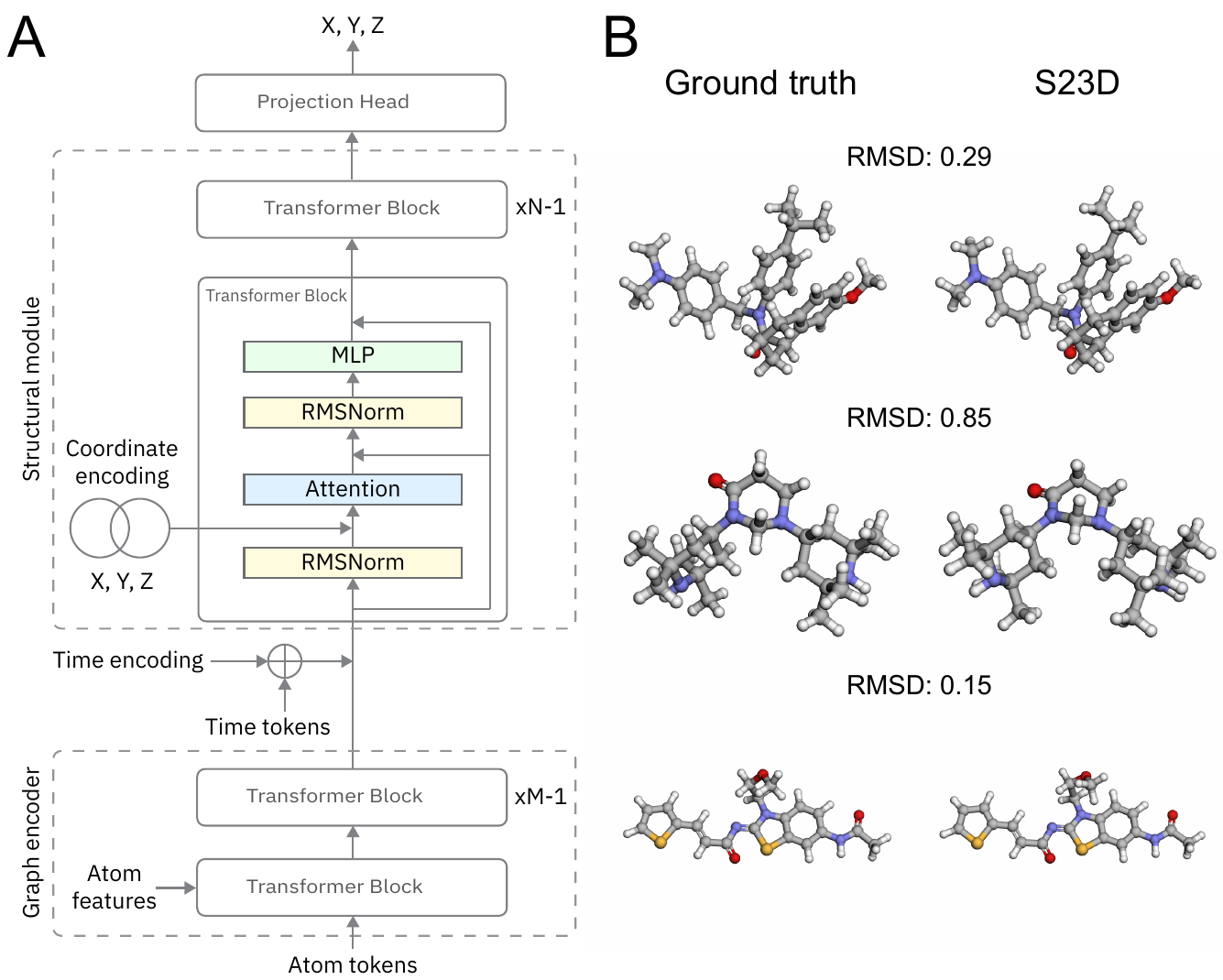}
  \caption{\textbf{A}: S23D score network architecture. The backbone of the model is a standard transformer network. The first set of transformer blocks (Graph encoder) encode the 2D graph and atom features, which are projected into hidden states in the first transformer block. Coordinate encodings are projected into hidden states of the first transformer block in the second set of blocks (structural module), where time tokens are also introduced. The head of the score network is a projection layer that transforms each atom embedding into $\mathbb{R}^{3}$. \textbf{B}: Example generated conformations. Ground truth conformations from GEOM-DRUGS test set (left) and aligned model generated samples (right).}
  \label{fig:Model structure}
\end{figure}

\section{Methods}
\label{sec:methods}

\subsection{Model architecture}

In this paper, we used a diffusion model (Variance Preserving SDE \cite{song2020score}, for equations see \ref{sec:DIFF}) with a transformer score network. The overall structure of the score network is shown in Figure \ref{fig:Model structure}. The score network is composed of two subnets. The first subnetwork encodes 2D graph structure and transforms atom element embeddings into graph atom representations. The output of this network does not depend on atom coordinates and diffusion time. The output of the graph encoder is fed to a structural subnetwork that also receives 3D coordinates as input, and produces the score. Both subnetworks incorporate graph bias in the attention mechanism, as discussed below. The introduction of the graph encoder allows faster inference at larger model sizes, as the graph encoder only needs to be run once during the inference stage, rather than on each diffusion step. The backbone of the model is the transformer block from the LLaMA architecture \cite{touvron2023llama}. The only modifications to the LLaMA transformer block were: 1) injection of atom features into hidden states after the first normalization layer; and 2) linear attention bias calculated from shortest path distances. The final layer of the network is a linear layer for projection of atom embeddings into $\mathbb{R}^3$. Below, we describe components of the augmented transformer architecture and underlying motivation in detail.

\subsubsection{2D molecular graph encoding}
\label{sec:2dmolgraphenc}

Input tokens for the transformer were atom tokens. For atom tokenization, we selected chemical elements with frequent occurrences (B, Bi, Br, C, Cl, F, H, I, N, O, P, S, Si) and assigned a unique token embedding for each. For all other chemical elements, the embedding of the special \textbf{MASK} token was used.

The LLaMA transformer block contains normalization, attention, second normalization, and MLP layers. In the graph encoder, molecular graph features $f_i$ for an atom $i$ were injected between the first normalization and attention layers by projecting features into atom hidden states. In addition, similar to \cite{ma2023graph, corso2020principal}, we introduced an adaptive degree scaler to inject graph node degree $deg_i$ into a hidden state $\mathbf{h}_i \in \mathbb{R}^{1\times d}$ for an atom $i$,
\begin{equation} \label{eq:DEG_SCALE}
\mathbf{h}^{out}_i=\mathbf{h}_i\odot\bm{\theta}_1+ \mathbf{h}_i\odot\bm{\theta}_2 \cdot log(1+deg_i) + \mathbf{f}_i\mathbf{W}_f,
\end{equation}
where $\bm{\theta}_1$ and $\bm{\theta}_2 \in \mathbb{R}^{1\times d}$ are learnable weights. The binary feature vector $\mathbf{f}_i\in \mathbb{R}^{1 \times 5}$ projected by the weight $\mathbf{W}_f \in \mathbb{R}^{5\times d}$ consists of bond type and atom charge. Three entries in the feature vector corresponds to distinct bond types (double, triple, aromatic), and a value of 1 means the atom is connected to at least one other atom with a bond of that type, while 0 means it is not. Two entries in $\mathbf{f}_i$ are allocated to a one-hot representation of positive, zero, and negative atom charges. We did not use chirality features as the main goal of the paper was to make a fair comparison between our and MCF models.

Implementation of the learnable bias in \eqref{eq:graphormer} requires querying lookup tables, which although highly optimized is still not an efficient operation on GPUs. Therefore, to incorporate graph structure, we used a modification of \eqref{eq:graphormer} in all multi-head attention layers,
\begin{equation} \label{eq:Alibi}
    A^{head}_{ij} = \frac{(\mathbf{h}_i\mathbf{W}^{head}_Q)(\mathbf{h}_j\mathbf{W}^{head}_K)^T}{\sqrt{d^{head}}}-m_{head}\phi(v_i, v_j),
\end{equation}
 where $head$ specifies the Query-Key product matrix $\mathbf{A}$ and weights $\mathbf{W}^{head}_Q$ and $\mathbf{W}^{head}_K$ for a specific attention head, and $m_{head}$ is a head-specific slope. We used slopes that were used in the ALiBi positional encoding \cite{press2021train}, a geometric sequence that starts with $2^{\frac{-8}{n}}$ for $n$ attention heads and has the same value for the ratio. Since we had a mixture of time tokens and atom tokens in the structural subnetwork, zero bias with atoms tokens was applied for time tokens.

\subsubsection{Spatial feature encoding}
\label{sec:spatialfeatureencoding}
For 2D graph features, atom $i$ coordinate encodings $\mathbf{s}_i$ were projected into hidden states of atoms after the first normalization layer in the first transformer block of the second subnetwork, 
\begin{equation} \label{eq:COORD_PROJ}
\mathbf{h}^{out}_i= \mathbf{h}_i + \mathbf{s}_i\mathbf{W}_c.
\end{equation}
In this work, we tested multiple types of continuous encoders modified from those proposed previously \cite{gorishniy2022embeddings} to obtain $\mathbf{s}_i$: 
\begin{enumerate}
\item A plain coordinate encoding, $\mathbf{s}_i = \mathbf{r}_i = [x_1, x_2, x_3]$, where $x_i$ are plain normalized atom coordinates.
\item A sinusoidal encoding combined with atom coordinates, here atom coordinates $r_i$ were complemented with their sinusoidal encoding, $[sin(w_{jc}x_c), cos(\bar{w}_{jc}x_c)]$, where $w_{j,c}$ and $\bar{w}_{jc}$ are trainable model weights initialized with $2^j\pi$, $j=1,\dots, 6$. 
\item An MLP encoding, $\mathbf{r}_i$ is transformed into $\mathbf{s}_i$ using two layer MLP with 128 hidden size and ReLU non-linearity.  
\end{enumerate}

\subsubsection{Data augmentation and chirality correction} \label{sec:ChirCor}
Random orthogonal transformations with matrices drawn from the O(3) Haar distribution (using class ortho\_group from SciPy) were applied to all conformations. Note that non-equivariant models trained on GEOM-DRUGS without chirality features produce samples with random chirality even if augmentation transformations are from SO(3) group, and we decided to use O(3) group augmentation with \textit{post hoc} chirality correction. Coordinate inputs to the transformer were translated by a random shift with components drawn from a uniform distribution with translations up to $30\text{\AA}$. We correct chirality with mirror reflection of conformations if Cahn-Ingold-Prelog labels produced by RDKit did not match the labels of the target molecule. Note, this operation corrects chirality of all conformers with single chiral centers, but rarely succeeds for conformers with multiple chiral centers.

\section{Experiments}
\label{sec:experiments}

We empirically evaluated our model by comparing generated conformers with reference conformers given an input SMILES. We used metrics based on the RMSD between generated and reference conformers, as described in \eqref{eq:RMSD}-\eqref{eq:AMR}. We evaluated the model using GEOM datasets, reporting results from GEOM-DRUGS as our primary experiment in section \ref{sec:geomdrugs}, with further breakdown of performance in the Appendix in section \ref{sec:geomdrugsappendix}. Also in the Appendix, we conduct ablation studies on positional encoding (\ref{sec:peablation}), report results of training without hydrogens (\ref{sec:onestagetraining}) report benchmark results on GEOM-QM9 and GEOM-XL (\ref{sec:geomqm9} and \ref{sec:geomxl}, respectively), compare inference sampling time with MCF (\ref{sec:samplingtime}), and report performance with varying numbers of inference steps (\ref{sec:inferencesteps}).

\subsection{Experimental Setup}

Our model, which we termed `S23D' (``SMILES to 3D''), was trained with two primary configurations of different sizes, `small' (S23D-S, 8.6M parameters) and `base' (S23D-B, 24.8M parameters). Parameters of these two primary models are described in section \ref{sec:TransPars}. We trained several variants of S23D-S with different options. We tested the effect of training with different numbers of transformer blocks for the graph encoder and structural subnetwork sections of the model, while maintaing the same total model size, which we labeled with `-M/N', respectively (see section \ref{sec:TransPars}, e.g. S23D-S-1/9 indicates $1$ graph encoder block and $9$ structural blocks). We also tested different coordinate encodings, which we label `M' (MLP), `S' (sinusoidal), or `C' (plain coordinates) (see section \ref{sec:spatialfeatureencoding}, e.g. S23D-S-1/9 (S) indicates sinusoidal encoding was used).

To enable direct comparison with MCF, we trained models on the GEOM-DRUGS dataset with splits as in \cite{ganea2021geomol}, which are also used by Torsional Diffusion \cite{jing2022torsional}, MCF \cite{wang2024swallowing} and ET-Flow \cite{hassan2024flow}. All experiments were conducted in two stages, first training for 100 epochs with hydrogen atoms removed, and second training for a further 25-35 epochs on the same data using complete molecules. MCF was trained for $750,000$ steps using batch size $512$, which is equivalent to $93.6$ epochs in our experiments. At the start of the second stage the embedding vector for hydrogen was initialized with the pre-trained embedding of the \textbf{MASK} token. During both training stages, we masked 10\% of atoms with the token \textbf{MASK} for 1\% of the molecules, selecting atoms to mask randomly with a probability inversely proportional to their frequencies in a given molecule.

Our primary comparisons were against MCF models, the current SOTA non-equivariant model, of similar size to our models (i.e. MCF-S and MCF-B), so we initially used 20 conformations per molecule and no chirality correction to match the experimental setup in \cite{wang2024swallowing}. To understand how the number of conformations per molecule affects model performance, we also trained with 30 conformations per molecule, as used in \cite{jing2022torsional, wang2024swallowing}, reducing the number of training epochs to 80 to offset the increased size of the training data. To provide a good comparison with ET-Flow, the current SOTA equivariant model, we trained with both 30 conformations per molecule and introduced an additional chirality correction (described in section \ref{sec:ChirCor}). Inference for all model variants was performed using 300 diffusion steps with a batch size of 128.

\begin{table*}[t]
\caption{Molecule conformer generation results on GEOM-DRUGS ($\delta = 0.75\text{\AA}$). Sections of the table are seperated by approximate size of the model. \textcolor{blue}{Blue} shows SOTA results within a model size range if the metric is only improved upon by a larger model, and \textbf{bold} shows overall SOTA results. `Size' is model size in millions of parameters. `$K_G$' is number of conformers for each molecule in test split used during training. `Chirality' indicates whether chirality information was used, either as features during training or corrected during generation. S23D-S and S23D-B indicate our small and base models respectively. `-M/N' show numbers of layers in each model section (see section \ref{sec:TransPars}). (M), (S), and (C) show the coordinate encoding scheme used: MLP, sinusoidal, or plain coordinates, respectively (see section \ref{sec:spatialfeatureencoding})}
\label{tab:geomdrugs}
\vskip 0.15in
\begin{center}
\begin{scriptsize}
\begin{sc}
\begin{tabular}{lccccccccccc}
\toprule
& & & & \multicolumn{4}{c}{Recall} & \multicolumn{4}{c}{Precision} \\
& & & & \multicolumn{2}{c}{Coverage $\uparrow$} & \multicolumn{2}{c}{AMR $\downarrow$} & \multicolumn{2}{c}{Coverage $\uparrow$} & \multicolumn{2}{c}{AMR $\downarrow$} \\
\midrule
& Size (M) & $K_G$ & Chiral. & mean & med. & mean & med. & mean & med. & mean & med. \\
\midrule
GeoMol &  & 20 & Yes & 44.6 & 41.4 & 0.875 & 0.834 & 43.0 & 36.4 & 0.928 & 0.841 \\
Tor. Diff. &  & 30 & Yes & 72.7 & 80.0 & 0.582 & 0.565 & 55.2 & 56.9 & 0.778 & 0.729 
\\
MCF-S & 13 & 20 & No & 79.4 & 87.5 & 0.512 & 0.492 & 57.4 & 57.6 & 0.761 & 0.715 \\
ET-Flow - SS & 8.3 & 30 & Yes & 79.6 & 84.6 & 0.439 & \textcolor{blue}{0.406} & \textbf{75.2} & \textbf{81.7} & \textbf{0.517} & \textbf{0.442} \\
ET-Flow - \it{SO(3)} & 9.1 & 30 & Yes & 78.2 & 83.3 & 0.480 & 0.459 & 67.3 & 71.2 & 0.637 & 0.567 \\
S23D-S-1/9 (S) & 8.6 & 20 & No & 80.7 & 89.9 & 0.483 & 0.458 & 57.5 & 57.5 & 0.757 & 0.700 \\
S23D-S-4/6 (S) & 8.6 & 20 & No & 81.5 & 89.4 & 0.475 & 0.455 & 59.6 & 61.4 & 0.732 & 0.677 \\
S23D-S-1/9 (M) & 8.6 & 20 & No & 81.5 & 90.0 & 0.473 & 0.444 & 58.1 & 57.8 & 0.751 & 0.702 \\
S23D-S-1/9 (C) & 8.6 & 20 & No & 81.9 & 88.9 & 0.462 & 0.442 & 58.8 & 58.4 & 0.740 & 0.697 \\
S23D-S-1/9 (S) & 8.6 & 30 & No & 81.5 & 89.6 & 0.475 & 0.451 & 58.4 & 58.9 & 0.749 & 0.692 \\
S23D-S-1/9 (S) & 8.6 & 20 & Yes & 83.3 & 91.9 & 0.457 & 0.432 & 63.4 & 67.0 & 0.689 & 0.630 \\
S23D-S-1/9 (M) & 8.6 & 20 & Yes & 83.8 & 91.7 & 0.449 & 0.419 & 63.8 & 66.7 & 0.682 & 0.628 \\
S23D-S-1/9 (C) & 8.6 & 20 & Yes & \textcolor{blue}{84.2} & 91.6 & \textcolor{blue}{0.438} & 0.420 & 64.8 & 68.2 & 0.668 & 0.607 \\
S23D-S-1/9 (S) & 8.6 & 30 & Yes & 83.8 & \textcolor{blue}{92.0} & 0.449 & 0.417 & 64.1 & 67.5 & 0.680 & 0.622 \\
\midrule
MCF-B & 64 & 20 & No & 84.0 & 91.5 & 0.427 & 0.402 & 64.0 & 66.8 & 0.667 & 0.605 \\
S23D-B-1/13 (S) & 25 & 20 & No & 84.5 & 91.7 & 0.420 & 0.387 & 62.5 & 63.4 & 0.690 & 0.626 \\
S23D-B-1/13 (M) & 25 & 20 & No & 84.6 & 91.9 & 0.412 & 0.381 & 62.4 & 64.1 & 0.684 & 0.613 \\
S23D-B-1/13 (C) & 25 & 20 & No & 84.2 & 91.2 & 0.419 & 0.388 & 62.9 & 65.0 & 0.683 & 0.621 \\
S23D-B-1/13 (S) & 25 & 20 & Yes & 86.5 & 93.6 & 0.391 & 0.363 & 69.2 & 74.6 & 0.608 & 0.539 \\
S23D-B-1/13 (M) & 25 & 20 & Yes & \textbf{87.0} & \textbf{94.0} & \textbf{0.380} & \textcolor{blue}{0.356} & 69.7 & 75.0 & 0.599 & 0.532 \\
S23D-B-1/13 (C) & 25 & 20 & Yes & 86.3 & 93.8 & 0.392 & 0.363 & 69.4 & 73.2 & 0.602 & 0.542 \\
\midrule
MCF-L & 242 & 20 & No & 84.7 & 92.2 & 0.390 & \textbf{0.247} & 66.8 & 71.3 & 0.618 & 0.530 \\
\bottomrule
\end{tabular}
\end{sc}
\end{scriptsize}
\end{center}
\vskip -0.1in
\end{table*}

\subsection{GEOM-DRUGS}
\label{sec:geomdrugs}

We report results on recreating conformations from the GEOM-DRUGS dataset \cite{axelrod2022geom} (CC0 1.0 license), compared with relevant baseline models, in Table \ref{tab:geomdrugs}. The dataset and benchmark are described in the dataset subsection of Section \ref{sec:relatedwork}. Table \ref{tab:geomdrugs} is divided into sections based on categories of approximate model size, small (8-13M parameters), medium (25-64M parameters) and large (only MCF-L at 242M parameters), with columns indicating the number of conformers per molecule used during training (`$K_G$') and whether chirality information was used during training or generation (`Chiral.'), to attempt to highlight the impact of various decisions made during model feature selection, data processing and training across previous models. 

Our first model configuration, S23D-S-1/9 (S), shows a substantial improvement in quality of generated conformers relative to MCF-S, indicated by an improvement in all RMSD metrics, despite being $\approx 33\%$ smaller, thereby establishing a new SOTA for a small, non-equivariant MCG model. Within S23D-S variants, we observed similar performance from all coordinate encoding methods tested. Plain coordinate (C) and MLP (M) encodings gave a $\approx 1\%$ improvement in mean coverage for both recall and precision, compared with sinusoidal (S) encoding, but other metrics, such as median recall coverage, were not improved, so we left sinusoidal encoding as the default for further variant testing. A small improvement in some metrics seen when using 30 conformers per molecule rather than 20. Altering the ratio of graph encoder to structural processing blocks from 1/9 to 4/6 resulted in similar recall metrics but showed a noticeable improvement in all precision metrics.

S23D-B, our `base' size model with $24.8$M parameters, shows improved quality of generated conformers relative to MCF-B in all recall RMSD metrics despite being 61\% smaller. In fact, S23D-B-1/13 (M) is within $0.1\%$ recall coverage of MCF-L, and close on other recall metrics. MCF-B performs better in precision metrics, but, as S23D-S improved on the precision metrics of MCF-S with less of a model size differential, we estimate that further scaling of the S23D-B model will surpass precision metric performance of MCF-B before reaching the same model size.

\subsubsection{Impact of chirality correction}

Small model SOTA for recall AMR and precision coverage and AMR is set by ET-Flow, which used both chirality information as features and a chirality correction operation on generated conformers, inverting the molecule along one axis if incorrect chirality was detected after generation \cite{hassan2024flow}. The ET-Flow - \textit{SO(3)} variant modified the architecture to improve chirality matching, but removed the chirality correction step, which resulted in lower performance on RMSD metrics than ET-Flow with chirality correction, but still surpassed MCF-S in recall AMR as well as precision metrics. S23D-S variants show greatly improved recall coverage relative to ET-Flow variants, and S23D-S-1/9 (S) shows improved recall AMR when trained with 30 conformations per molecule, despite not using any chirality information during training or generation. Applying our simple chirality correction to the results from our S23D-S-1/9 (S) model trained with 20 or 30 conformers produced a $>2\%$ increase in recall coverage and $\approx 6\%$ increase in precision coverage metrics.  However, the equivariant ET-Flow model remains SOTA for precision, whether using chirality features or chirality correction. Note that this model even vastly outperforms the 242M parameter MCF-L model on precision metrics, and used additional augmentations such as a harmonic prior and rotational alignment at the start of coordinate generation. Nevertheless, our best `small' model produced a new SOTA for recall coverage, with mean $84.2\%$ and median $91.6\%$, for any model previously trained under $64$M parameters.

When applying simple chirality correction to our `base' size model, our S23D-B-1/13 (M) variant achieved a new SOTA for recall performance on GEOM-DRUGS, with mean coverage of $87.0\%$ and mean AMR of $0.380$, outperforming the MCF-L results with a model $\approx 10\%$ of the size. Precision coverage and mean AMR are also improved over MCF-L and only behind one set of results, the ET-Flow variant that used both chirality features and a chirality correction step.

\section{Limitations}
\label{sec:limitations}

In the current model we did not use chirality information to construct the model features, which we view as the main limitation of this work and the immediate next direction of our research. We anticipate that providing the model the information required to understand stereochemistry will lead to a quantitative leap in accuracy and efficiency of MCG. 
Current splits used in the GEOM benchmarks are random and not based on molecule scaffolds, which could lead to limited real-world application even for models that achieve SOTA on these benchmarks \cite{guo2024scaffold}. We have prepared new splits for our next study using Butina clastering \cite{hernandez2023nci60} on Murcko scaffolds \cite{wu2018moleculenet} to test model generalization ability. Additionally, to speed up MCG for practical applications it is critical to explore how model quantization and distillation would affect the current results.

\section{Conclusions}
\label{sec:conclusions}
In this paper, we advanced architectures of diffusion models for MCG, introducing a new transformer model based on a standard transformer architecture combined with a new graph PE. The PE is represented as attention with a bias term that is linear with respect to the shortest path distance between atoms in a molecular graph. Our model demonstrates superior performance across multiple model sizes on a model-to-model basis. Two-stage training on molecules without hydrogen representation, followed by finetuning on complete molecules, will facilitate training of larger models on large datasets within a limited computational budget. This transformer can become the basis of large foundation models for accurate prediction of molecular structure. Our work has potential impact in key areas of molecular modeling and supports AI as a valuable tool across pharmaceutical research, materials science, and computational chemistry. Potential societal implications of this research arise from the use of AI tools in the advancement of drug discovery and the field of chemistry in general. Ethical concerns about this direction may include a lack of transparency in how molecule properties are calculated and the potential for misunderstanding of the results generated by AI models.

\section*{Acknowledgments}
We would like to thank Joseph Morrone and Parthasarathy Suryanarayanan for insightful discussions.

\bibliographystyle{plain}
\bibliography{main}







\newpage
\appendix

\section{Model details}
\label{sec:AppModelDetails}
\subsection{Model and training parameters} \label{sec:TransPars}
We trained the model with Adam optimizer with default PyTorch parameters and a constant learning rate 0.001, weight decay of 0.001 and batch size of 128. We trained the model on 2 V100 GPUs. GTX1660 and Quadro M5000 GPUs were used for inference.

\begin{table}[h]
    \centering
    \caption{Transformer parameters}
    \begin{tabular}{|l|c|c|c|c|}
        \hline
        \textbf{Parameter} & \textbf{S23D-S-1/9} & \textbf{S23D-S-4/6} & \textbf{S23D-B-1/13}\\
        \hline
        Number of graph encoder blocks & 1 & 4 & 1  \\
        Number of blocks in structure subnetwork & 9 & 6 & 13  \\
        Hidden Size &  256 & 256 & 384 \\
        Number of Attention Heads & 8 & 8 & 12 \\
        \hline
    \end{tabular}
    \label{tab:transformer_comparison}
\end{table}

\subsection{Diffusion model} \label{sec:DIFF}

Following \cite{song2020score}, we used stochastic differential equation (SDE)
\begin{equation}
    d\bm{x} = f(t)\bm{x}dt+g(t)d\bm{w},
\label{eq:sde}
\end{equation}
with initial condition $x(0) \sim p_0$ and $t\in(0,1)$, where $\bm{w}$ is a standard Wiener process, term $f(t)\bm{x}$ is a drift term and $g(t)$ is a diffusion term. Here $\bm{x}$ is a vector of stacked atom coordinates normalized by the constant $20\text{\AA}$. For variance preserving SDE (VPSDE), $f(t) = -\frac{\beta(t)}{2}$, $g(t) = \sqrt{\beta(t)}$, and $p(\bm{x}(t)| \bm{x}(0)) = \mathcal{N}(\bm{x}(t);\bm{x}(0)e^{-\frac{1}{2}\gamma(t)},\bm{I}-\bm{I}e^{-\gamma(t)})$, where $\gamma(t) = \int_{0}^{t} \beta(s)\,ds$. We used linear $\beta(t)=(\beta_e-\beta_s)t + \beta_s$ with $\beta$ in $(0.0, 18.0)$. 
\begin{equation}
\begin{aligned}
\gamma(t) =& \int_{0}^{t} \left[(\beta_e-\beta_s) t + \beta_s\right]ds= \\
=&(\beta_e-\beta_s) \frac{t^2}{2} + \beta_s t
\end{aligned}
\end{equation}

The general formula to obtain noised samples in diffusion models is $\bm{x}(t) = \alpha (t) \bm{x}(0) + \sigma(t) \bm{\eta}$, with $\alpha(t) = e^{-\frac{1}{2}\gamma(t)}$ and $\sigma(t) = \sqrt{1-e^{-\gamma(t)}}$ for VPSDE. The noise schedule is highly important for the final model performance, as in \cite{chen2023importance} we modified the noise schedule  
\begin{equation}
    \bm{x}(t) = \alpha (t)b \bm{x}(0) + \sigma(t) \bm{\eta},
\end{equation}
effectively rescaling atom coordinates by the constant $b=20$, and the input to the score network was scaled by $b_s(t) = \frac{1}{\sqrt{(b^2-1)e^{-\gamma(t)}+1}}$ to keep stable network input variance. To train the score transformer $s_\theta(x, G, t)$, we used loss
\begin{equation} \label{eq:noiserec}
    \bm{\theta^*}=\operatorname*{arg\, min}_{\bm{\theta}}\E_t\left[\E_{G, \bm{x}(0)}\E_{\bm{x}(t)|G, \bm{x}(0)}\E_{\bm{\eta}}\frac{1}{|V(G)|}||s_\theta(x(t)\cdot b_s(t), G, t)\cdot\sigma(t)+\bm{\eta}||_2^2\right],
\end{equation}
 where $G$ is a 2D molecular graph with number of atoms $|V(G)|$. The Euler–Maruyama method for reversed process was used to sample atom coordinates. 

 For molecules with size $M$, we worked in subspaces $X=\{\bm{x} \in \mathbb{R}^{M\times3}: \frac{1}{M}\sum_{i=1}^{M}\bm{x}_i=\bm{0}\}$ by subtracting center of mass (CoM) from noise after sampling as shown in \cite{bao2022equivariant}. We also subtracted the mean value from the output of the score network. 

 Multiple time tokens, each with a different embedding, were inputs to the score network. We used 4 time tokens instead of a single token to exploit the benefits of memory tokens \cite{burtsev2020memory}. The projection of diffusion time encoding $TE$ was added to the time token embeddings, 
\begin{equation} \label{eq:time encoding}
\begin{aligned}
TE(t, 2i) &=& sin(2^i\pi t) \\
TE(t, 2i + 1) &=& cos(2^i\pi t) 
\end{aligned}
\end{equation}
where $i=1,\dots, 10$. 

\section{Additional experiments}

\subsection{Ablations}

\subsubsection{Positional encoding ablation}
\label{sec:peablation}

\begin{table*}[t]
\caption{Positional encoding ablation results on GEOM-DRUGS ($\delta = 0.75\text{\AA}$). The model used is S23D-S-1/9 (S) and `ALiBi' PE represents the PE scheme with fixed bias with learnable slope that we present here (and reported in Table \ref{tab:geomdrugs}. `Eigenvectors' and `Learnable bias' represent the PE schemes used in MCF \cite{wang2024swallowing} and Graphormer \cite{ying2021transformers}, respectively (see section \ref{sec:introduction})}
\label{tab:peablation}
\vskip 0.15in
\begin{center}
\begin{scriptsize}
\begin{sc}
\begin{tabular}{lcccccccccc}
\toprule
& & & \multicolumn{4}{c}{Recall} & \multicolumn{4}{c}{Precision} \\
& & & \multicolumn{2}{c}{Coverage $\uparrow$} & \multicolumn{2}{c}{AMR $\downarrow$} & \multicolumn{2}{c}{Coverage $\uparrow$} & \multicolumn{2}{c}{AMR $\downarrow$} \\
\midrule
& PE & Epochs & mean & med. & mean & med. & mean & med. & mean & med. \\
\midrule
S23D-S-1/9 (S) & ALiBi & 100 & 81.4 & 89.0 & 0.472 & 0.449 & 57.5 & 57.0 & 0.756 & 0.709 \\
S23D-S-1/9 (S) & Eigenvectors & 100 & 79.5 & 86.9 & 0.507 & 0.479 & 55.6 & 54.5 & 0.789 & 0.739 \\
S23D-S-1/9 (S) & Learnable bias & 100 & 81.5 & 88.9 & 0.468 & 0.441 & 58.6 & 59.5 & 0.744 & 0.693 \\
\midrule
S23D-B-1/13 (S) & Eigenvectors & 85 & 82.7 & 90.0 & 0.449 & 0.422 & 60.0 & 60.1 & 0.725 & 0.664 \\
\bottomrule
\end{tabular}
\end{sc}
\end{scriptsize}
\end{center}
\vskip -0.1in
\end{table*}

To assess the impact of choice of PE on model performance, we trained 3 versions of the S23D-S-1/9 (S) model using a different PE in each. The three PEs used were `ALiBi', our default PE described in section \ref{sec:2dmolgraphenc}, `Eigenvectors', the PE scheme used in MCF \cite{wang2024swallowing}, and `Learnable bias', the PE scheme used in Graphormer \cite{ying2021transformers}. The `Eigenvectors' PE uses the $k$ eigenvectors of the graph Laplacian with largest eigenvalues to encode the graph nodes. We used $k=28$ and used the eigenvectors pre-computed for the GEOM-DRUGS dataset directly provided by \cite{wang2024swallowing}. The `Learnable bias' PE used equation \ref{eq:graphormer} updated for multiple heads, as in Graphormer (with maximum shortest path threshold of 20, as in the default Graphormer settings) \cite{ying2021transformers}. We trained PE ablations using a single stage of training with hydrogens due to the fact that in the `Eigenvectors' PE the eigenvectors were pre-computed with hydrogens included, so we were unable to follow our two-stage training approach. We trained for 100 epochs, which amounts to more training than in our two-stage approach, where we trained for 100 epochs without hydrogens (approximately two times faster than training with hydrogens) and then fine-tuned with hydrogens for 35 epochs (equivalent to a total of approximately 85 epochs of single-stage training). PE ablation results are shown in Table \ref{tab:peablation}. Using the eigenvector positional encoding produced results that are consistent with the MCF-S model (see Table \ref{tab:geomdrugs}), that are worse than the results using the `ALiBi' or `Learnable Bias' PE schemes in all metrics. To further check whether the performance of the eigenvectors PE would scale more effectively than `ALiBi' or `Learnable Bias', we trained the S23D-B-1/13 (S) model for 85 epochs (single stage, no hydrogens, no chirality correction) with eigenvectors PE and found mean recall coverage $\approx 1.9\%$ worse than the results in Table \ref{tab:geomdrugs} obtained using `ALiBi', the same performance deficit as we found for the `Eigenvectors' PE in the smaller S23D-S model. The `Learnable Bias' performed similarly to `ALiBi', but training was $\approx35\%$ slower than `ALiBi' on A100 GPUs due to inefficient lookup operations, as discussed in section \ref{sec:2dmolgraphenc}. We therefore maintained the use of the `ALiBi' PE for all other experiments.

\subsubsection{Training without hydrogen}
\label{sec:onestagetraining}

After the first stage of training in our two-stage training approach the model generates heavy atom coordinates without hydrogen atom coordinates. However, it is possible to use fast, rule-based approaches to add hydrogens at this stage, and assess the performance of the model. Using the `Chem.AddHs' function in RDKit, we evaluated our models after the first stage of training. Results on GEOM-DRUGS are shown in Table \ref{tab:nohydrogens}. After the first stage of training without hydrogens, and with no chirality correction, mean recall coverage values for our `small' model variants ($80.3-81.0\%$) surpassed the previous SOTA for small models (ET-Flow - SS: $79.6\%$), and mean recall coverage values for our `base' model variants ($84.0-84.2\%$) surpassed SOTA for the `base' model size (MCF-B: $84.0\%$), showing that training without hydrogen atoms is a viable approach. Our primary motivation for the second stage of training was for a direct comparison with previous methods, which all generate coordinates for all atoms, including hydrogens. After the second stage of training mean recall coverage improved by approximately $0.5\%$ for `small' and `base' model variants (see Table \ref{tab:geomdrugs}). 

\begin{table}[h]
\caption{Molecule conformer generation results generating heavy atoms only, and using rule-based approach to add hydrogens. Training for $100$ epochs for all model variants. Columns described in Table \ref{tab:geomdrugs}}
\label{tab:nohydrogens}
\vskip 0.15in
\begin{center}
\begin{scriptsize}
\begin{sc}
\begin{tabular}{lccccccccccc}
\toprule
& & & & \multicolumn{4}{c}{Recall} & \multicolumn{4}{c}{Precision} \\
& & & & \multicolumn{2}{c}{Coverage $\uparrow$} & \multicolumn{2}{c}{AMR $\downarrow$} & \multicolumn{2}{c}{Coverage $\uparrow$} & \multicolumn{2}{c}{AMR $\downarrow$} \\
\midrule
& Size (M) & $K_G$ & Chiral. & mean & med. & mean & med. & mean & med. & mean & med. \\
\midrule

S23D-S-1/9 (S) & 8.6 & 20 & No & 80.3 & 89.6 & 0.495 & 0.467 & 56.1 & 55.4 & 0.779 & 0.724 \\
S23D-S-1/9 (M) & 8.6 & 20 & No & 81.0 & 89.4 & 0.485 & 0.457 & 58.2 & 59.1 & 0.751 & 0.702 \\
S23D-S-1/9 (C) & 8.6 & 20 & No & 80.8 & 88.7 & 0.485 & 0.454 & 57.5 & 57.1 & 0.763 & 0.707 \\

\midrule

S23D-B-1/13 (S) & 25 & 20 & No & 84.1 & 91.2 & 0.428 & 0.399 & 62.2 & 64.3 & 0.694 & 0.625 \\
S23D-B-1/13 (M) & 25 & 20 & No & 84.2 & 91.9 & 0.425 & 0.391 & 61.7 & 63.2 & 0.701 & 0.632 \\
S23D-B-1/13 (C) & 25 & 20 & No & 84.0 & 91.5 & 0.433 & 0.401 & 62.5 & 64.4 & 0.690 & 0.632 \\

\bottomrule
\end{tabular}
\end{sc}
\end{scriptsize}
\end{center}
\vskip -0.1in
\end{table}

\subsection{GEOM datasets}

\subsubsection{GEOM-QM9}
\label{sec:geomqm9}

We trained the model on the GEOM-QM9 dataset \cite{axelrod2022geom}, introduced in the \textbf{Datasets and benchmarks for conformation prediction} section in section \ref{sec:relatedwork}, using the data processed and provided by \cite{wang2024swallowing}, see \href{https://github.com/apple/ml-mcf}{https://github.com/apple/ml-mcf}. The processed test dataset contained 995 molecules rather than the 1000 indexed in the test set of \cite{ganea2021geomol}, so we used 995 for comparison with the MCF model, assuming that they tested on the molecules provided in their processed dataset. We trained for $430$ epochs with batch size $512$. Inference was performed with $300$ diffusion steps, as in the GEOM-DRUGS experiments. In our data augmentation strategy for QM9 data, we switched from orthogonal transformations from O(3) group to rotations from SO(3) due to bias in the distribution of chirality in the QM9 dataset. Results are shown in Table \ref{tab:geomqm9}. S23D-B coverage metrics are comparable with MCF-B and ET-Flow models. Mean AMR values across these models are $<0.1$, which is around $10\%$ of the typical bond length between atoms, indicating that benchmark results on the QM9 metric are likely to be saturated. Therefore, we did not perform additional experimentation to improve S23D performance, and the QM9 benchmark may no longer be appropriate for testing SOTA MCG models.

\begin{table}[h]
\caption{Molecule conformer generation results on GEOM-QM9 ($\delta = 0.5\text{\AA}$). Columns described in Table \ref{tab:geomdrugs}}
\label{tab:geomqm9}
\vskip 0.15in
\begin{center}
\begin{scriptsize}
\begin{sc}
\begin{tabular}{lccccccccccc}
\toprule
& & & & \multicolumn{4}{c}{Recall} & \multicolumn{4}{c}{Precision} \\
& & & & \multicolumn{2}{c}{Coverage $\uparrow$} & \multicolumn{2}{c}{AMR $\downarrow$} & \multicolumn{2}{c}{Coverage $\uparrow$} & \multicolumn{2}{c}{AMR $\downarrow$} \\
\midrule
& Size (M) & $K_G$ & Chiral. & mean & med. & mean & med. & mean & med. & mean & med. \\
\midrule
GeoMol &  & 10 & Yes & 91.5 & \textbf{100.0} & 0.225 & 0.193 & 87.6 & \textbf{100.0} & 0.270 & 0.241 \\
Tor. Diff. &  & 30 & Yes & 92.8 & \textbf{100.0} & 0.178 & 0.147 & 92.7 & \textbf{100.0} & 0.221 & 0.195 \\
MCF-B & 64 & 10 & No & 95.0 & \textbf{100.0} & 0.103 & 0.044 & 93.7 & \textbf{100.0} & 0.119 & 0.055 \\
ET-Flow & 8.3 & 30 & Yes & \textbf{96.5} & \textbf{100.0} & \textbf{0.073} & 0.047 & \textbf{94.1} & \textbf{100.0} & \textbf{0.098} & \textbf{0.039} \\
ET-Flow - \it{SO(3)} & 9.1 & 30 & Yes & 96.0 & \textbf{100.0} & 0.076 & \textbf{0.030} & 92.1 & \textbf{100.0} & 0.110 & 0.047 \\
\midrule
S23D-B-1/13 (S) & 24.8 & 30 & No & 95.6 & \textbf{100.0} & 0.097 & 0.048 & 91.2 & \textbf{100.0} & 0.145 & 0.070 \\
S23D-B-1/13 (S) & 24.8 & 30 & Yes & 96.0 & \textbf{100.0} & 0.090 & 0.047 & 93.8 & \textbf{100.0} & 0.111 & 0.059 \\
S23D-B-1/13 (C) & 24.8 & 30 & No & 95.6 & \textbf{100.0} & 0.100 & 0.050 & 90.8 & \textbf{100.0} & 0.148 & 0.077 \\
S23D-B-1/13 (C) & 24.8 & 30 & Yes & 96.2 & \textbf{100.0} & 0.090 & 0.050 & 93.5 & \textbf{100.0} & 0.116 & 0.064 \\
\bottomrule
\end{tabular}
\end{sc}
\end{scriptsize}
\end{center}
\vskip -0.1in
\end{table}

\subsubsection{GEOM-XL}
\label{sec:geomxl}
As in other MCG papers \cite{jing2022torsional,wang2024swallowing,hassan2024flow}, we report results on GEOM-XL dataset, a subset of GEOM-MoleculeNet, to demonstrate how the model can generalize on large unseen molecules. For S23D-B-1/13 (S), we observed results comparable with MCF-B.

\begin{table}[h]
\caption{Molecule conformer generation results on GEOM-XL. Columns described in Table \ref{tab:geomdrugs}}
\label{tab:geomxl}
\vskip 0.15in
\begin{center}
\begin{scriptsize}
\begin{sc}
\begin{tabular}{lccccccc}
\toprule
& & & & \multicolumn{2}{c}{Recall} & \multicolumn{2}{c}{Precision} \\
& & & & \multicolumn{2}{c}{AMR $\downarrow$} & \multicolumn{2}{c}{AMR $\downarrow$} \\
\midrule
& Size(M) & $K_G$ & Chirality & mean & median & mean & median \\
\midrule
GeoMol &  & 20 & Yes & 2.47 & 2.39 & 3.30 & 3.14  \\
Tor. Diff. &  & 30 & Yes & 2.05 & 1.86 & \textbf{2.94} & 2.78  \\
MCF-S & 13 & 20 & No & 2.22 & 1.97 & 3.17 & 2.81  \\
MCF-B & 64 & 20 & No & 2.01 & 1.70 & 3.03 & 2.64  \\
MCF-L & 242 & 20 & No & \textbf{1.97} & \textbf{1.60} & \textbf{2.94} & \textbf{2.43}  \\
ET-Flow & 8.3 & 30 & Yes & 2.31 & 1.93 & 3.31 & 2.84  \\
S23D-B-1/13 (S) & 25 & 20 & No & 2.07 & 1.80 & 3.22 & 2.83 \\
\bottomrule
\end{tabular}
\end{sc}
\end{scriptsize}
\end{center}
\vskip -0.1in
\end{table}

\subsubsection{GEOM-DRUGS}
\label{sec:geomdrugsappendix}

\begin{figure}[h]
  \centering
  \includegraphics[width=\linewidth]{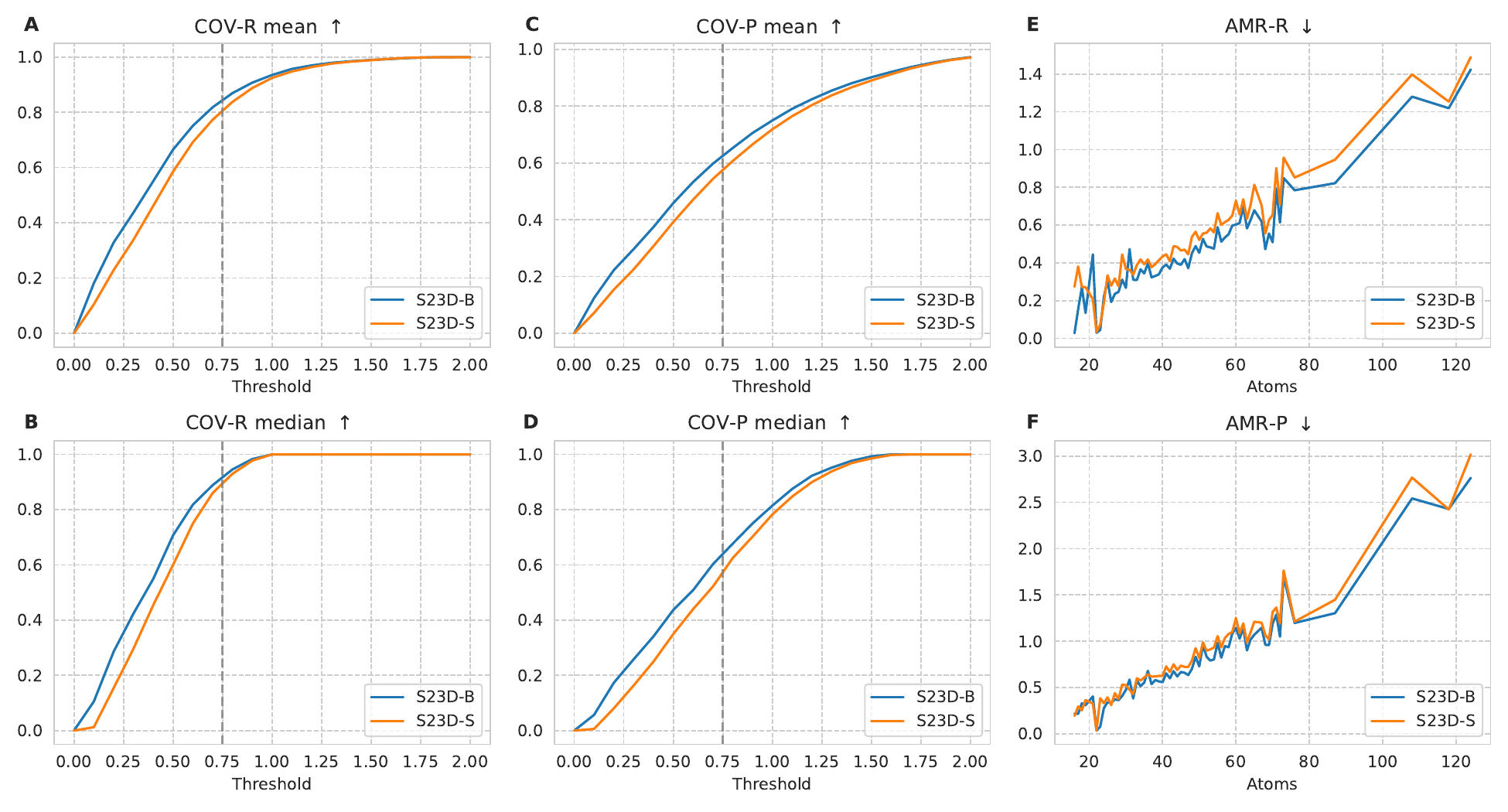}
  \caption{(\textbf{A-B}) Recall and (\textbf{C-D}) precision coverage as a function of coverage threshold, and \textbf{E} recall and \textbf{F} precision AMR as a function of number of atoms in a test molecule, for S23D-S-1/9 (S) and S23D-B-1/13 (S) (without chirality correction).}
  \label{fig:rmsdbreakdown}
\end{figure}

To understand performance of our S23D-S-1/9 (S) and S23D-B-1/13 (S) model on the GEOM-DRUGS test set in more detail, we examined the RMSD metrics for recall and precision as a function of the coverage threshold (Fig. \ref{fig:rmsdbreakdown}A-D) and number of atoms in the test molecules (Fig. \ref{fig:rmsdbreakdown}E-F). This breakdown can be compared with Figure 4 in \cite{wang2024swallowing} and Figure 3 in \cite{hassan2024flow} to examine performance of S23D models against Torsional Diffusion, MCF and ET-Flow.

\subsection{Inference sampling} \label{sec:InfSample}

\subsubsection{Sampling time}
\label{sec:samplingtime}

\begin{figure}[h]
  \centering
  \includegraphics[width=.8\linewidth]{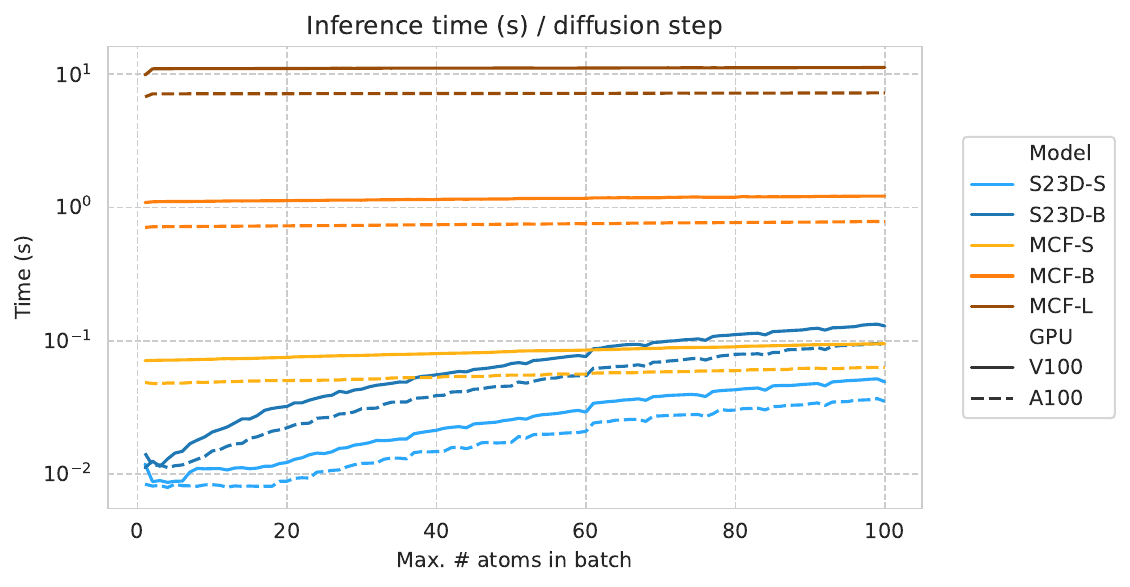}
  \caption{Inference time per diffusion step for S23D-S-1/9 (M), S23D-B-1/9 (M), MCF-S, MCF-B and MCF-L models on V100 and A100 GPUs over maximum number of atoms in a batch of size 256.}
  \label{fig:timing}
\end{figure}

To test inference speed, we measured time per step during the diffusion process. We created dummy batches of random data using different numbers of maximum atoms in each batch, and ran the diffusion process on both V100 and A100 GPUs. We averaged time over 1,000 diffusion steps for S23D-S, S23D-B and MCF-S, 200 steps for MCF-B, 20 steps for MCF-L on A100, and 10 steps for MCF-L on V100. We performed these computations on a shared cluster, so performance could be impacted by other CPU usage from other users on a cluster node. MCF models effectively double the batch size during inference to produce 2 conformations per input conformation, so we used batch size 128 in MCF models and 256 in S23D. Results are shown in Fig. \ref{fig:timing}. On a V100 GPU, S23D-S is $\approx4$ times faster per inference step than MCF-S for smaller molecules, and $\approx2$ times faster for larger molecules. On an A100 GPU, S23D-S is $\approx3$ times faster per inference step than MCF-S for smaller molecules, and $\approx1.5$ times faster for larger molecules.  S23D-B is $\approx2$ orders of magnitude faster than MCF-B for smaller molecules, and $\approx1$ order of magnitude faster than MCF-B for larger molecules, on either GPU model. MCF-L, which is outperformed on GEOM-DRUGS recall coverage metrics by S23D-B with chirality correction, is $\approx2$ orders of magnitude slower than S23D-B even for larger molecules.

Our architecture uses less computations than PerceiverIO (used in MCF), and is faster on devices preceeding the Ampere architecture, such as the V100, that do not support Flash Attention. MCF-S used a latent array in PerceiverIO of size 128, which is greater than the number of atoms in most molecules in the GEOM-DRUGS dataset. Thus, the latent layers of PerceiverIO have to process more ``tokens'' than our model, but this effect is reduced as molecule size approaches the latent size of 128. Currently Flash Attention doesn't support custom masks, so for S23D-S we are unable to benefit from the speed increase provided by Flash Attention on chips like the A100. In contrast, MCF doesn't require any mask in the latent layers and benefits from the Flash Attention optimization. 

It is important to note that the mean number of atoms per molecule in GEOM-DRUGS is $44.4$ \cite{axelrod2022geom}, with s.d. $11.3$, so MCF-S performance only approaches S23D-S performance on A100 for very rare batches in the dataset. Also, note that MCF performance metrics are reported using 1,000 inference steps, compared with 300 for S23D, making S23D-S $\approx 9$ times faster than MCF-S at the average molecule size to obtain the superior metrics reported in Table \ref{tab:geomdrugs}. We did not run timing comparisons against ET-Flow, but Figure 4 in \cite{hassan2024flow} demonstrates that ET-Flow is substantially slower per inference step than both MCF-S and MCF-B, which S23D models outperform.

\subsubsection{Inference steps}
\label{sec:inferencesteps}

We tested the interaction of different numbers of iterations during inference with the choice of encoding (sinusoidal, MLP, or projection) to understand whether any encoding scheme has a scaling benefit by outperforming at fewer inference steps. We tested with the S23D-S-1/13 model and (S), (M) and (C) variants using the Euler-Maruyama method at 50, 100, 200 and 300 inference steps. No chirality corrections were used. Results are shown in Table \ref{tab:Sampling}. Performance is similar at 300, 200, or 100 inference steps, and drops by a similar amount for all metrics across all model variants at 50 steps. We report report results in the main manuscript using 300 steps, but similar results could have been obtained using 100-200 steps.

We also tested performance scaling with number of inference steps using the larger S23D-B-1/13 (S) model with DDIM sampling \cite{song2020denoising}, to try and achieve a greater reduction in the number of steps required for this larger and slower model. Results are shown in Table \ref{tab:Sampling} using between 5 and 50 steps, compared with 300 in our default implementation. No chirality corrections were used. We observed a modest drop in metrics, when the number of sampling iterations with DDIM were reduced to 50 and 20 iterations. However, when compared with the results in Table \ref{tab:geomdrugs}, S23D-B outperforms MCF-S on recall metrics using only $10$ iteration steps with DDIM.

\begin{table}[h] 
\caption{Molecule conformer generation results on GEOM-DRUGS ($\delta = 0.75\text{\AA}$) using different numbers of inference steps during sampling. Method shows inference method. Steps shows number of inference steps.}
\label{tab:Sampling}
\label{tab:ddim}
\vskip 0.15in
\begin{center}
\begin{scriptsize}
\begin{sc}
\begin{tabular}{lccccccccccc}
\toprule
& & & & \multicolumn{4}{c}{Recall} & \multicolumn{4}{c}{Precision} \\
& & & & \multicolumn{2}{c}{Coverage $\uparrow$} & \multicolumn{2}{c}{AMR $\downarrow$} & \multicolumn{2}{c}{Coverage $\uparrow$} & \multicolumn{2}{c}{AMR $\downarrow$} \\
\midrule
& Method & Steps & mean & med. & mean & med. & mean & med. & mean & med. \\
\midrule

S23D-S-1/13 (S) & Euler–Maruyama & 300 & 80.7 & 89.9 & 0.483 & 0.458 & 57.5 & 57.5 & 0.757 & 0.700 \\
S23D-S-1/13 (S) & Euler–Maruyama & 200 &  81.1 & 89.8 & 0.483 & 0.455 & 57.7 & 57.1 & 0.755 & 0.701 \\
S23D-S-1/13 (S) & Euler–Maruyama & 100 & 80.9 & 88.9 & 0.495 & 0.470 & 57.4 & 56.7 & 0.767 & 0.720 \\
S23D-S-1/13 (S) & Euler–Maruyama & 50 & 78.9 & 87.1 & 0.551 & 0.517 & 54.8 & 53.9 & 0.817 & 0.765 \\

\midrule

S23D-S-1/13 (M) & Euler–Maruyama & 300 & 81.5 & 90.0 & 0.473 & 0.444 & 58.1 & 57.8 & 0.751 & 0.702 \\
S23D-S-1/13 (M) & Euler–Maruyama & 200 & 81.7 & 89.5 & 0.476 & 0.451 & 57.6 & 56.9 & 0.753 & 0.706 \\
S23D-S-1/13 (M) & Euler–Maruyama & 100 & 81.4 & 89.3 & 0.488 & 0.461 & 57.2 & 56.3 & 0.764 & 0.713 \\
S23D-S-1/13 (M) & Euler–Maruyama & 50 & 79.2 & 87.1 & 0.546 & 0.518 & 54.9 & 53.1 & 0.817 & 0.758 \\

\midrule

S23D-S-1/13 (C) & Euler–Maruyama & 300 & 81.9 & 88.9 & 0.462 & 0.442 & 58.8 & 58.4 & 0.740 & 0.697 \\
S23D-S-1/13 (C) & Euler–Maruyama & 200 & 81.2 & 88.3 & 0.474 & 0.445 & 57.8 & 57.2 & 0.753 & 0.703 \\
S23D-S-1/13 (C) & Euler–Maruyama & 100 & 81.5 & 88.9 & 0.485 & 0.457 & 58.2 & 57.0 & 0.759 & 0.702 \\
S23D-S-1/13 (C) & Euler–Maruyama & 50 & 79.2 & 86.7 & 0.546 & 0.517 & 55.6 & 54.6 & 0.809 & 0.743 \\

\midrule

S23D-B-1/13 (S) & Euler–Maruyama & 300 & 84.5 & 91.7 & 0.420 & 0.387 & 62.5 & 63.4 & 0.690 & 0.626 \\
S23D-B-1/13 (S) & DDIM & 50 & 83.0 & 91.3 & 0.447 & 0.422 & 58.8 & 59.4 & 0.734 & 0.676 \\
S23D-B-1/13 (S) & DDIM & 20 & 82.4 & 90.6 & 0.456 & 0.424 & 57.4 & 56.9 & 0.752 & 0.700 \\
S23D-B-1/13 (S) & DDIM & 10 & 81.0 & 89.7 & 0.477 & 0.450 & 55.5 & 54.9 & 0.776 & 0.723 \\
S23D-B-1/13 (S) & DDIM & 5 & 76.6 & 84.8 & 0.530 & 0.508 & 51.2 & 50.0 & 0.832 & 0.784 \\

\bottomrule
\end{tabular}
\end{sc}
\end{scriptsize}
\end{center}
\vskip -0.1in
\end{table}

\subsection{Conformation evaluation} \label{sec:confeval}
\subsubsection{PoseBusters}
\label{sec:posebusters}

To study the effect of coordinate encodings on generated molecular structure, we ran additional tests from the PoseBusters package \cite{buttenschoen2024posebusters}, including assessment of bond lengths, bond angles, aromatic ring flatness, planar double bonds, and internal steric clashes, on our generated molecules. Using the default criteria for "intramolecular validity" features in PoseBusters that a predicted feature (e.g. bond length) should be within $\pm25\%$ of the reference for that feature, all 5 of the above features were perfectly captured, with $>99.2\%$ pass rate for all model variants that we reported in Table 1. Results are shown in Table \ref{tab:posebusters}. Lowering the thresholds from $25\%$ to $10\%$ and $5\%$ we saw metrics drop evenly across model variants. For instance, at $5\%$ correct bond angles were found at a rate of $58.6\%$ for S23D-S-1/9 (S), $56.4\%$ for S23D-S-1/9 (M), and $54.8\%$ for S23D-S-1/9 (C).

\begin{table}[h] 
\caption{Molecule conformer validation results using PoseBusters measures of ``intramolecular validity''. Metric is \% of feature values from all generated conformations within $\pm$ threshold of target value for that feature. Results generated with $300$ inference steps and no chirality correction for full GEOM-DRUGS test set.}
\label{tab:posebusters}
\label{tab:ddim}
\vskip 0.15in
\begin{center}
\begin{scriptsize}
\begin{sc}
\begin{tabular}{lccccc}
\toprule
Feature & Threshold & S23D-S-1/9 (S) & S23D-S-1/9 (M) & S23D-S-1/9 (C) & S23D-B-1/13 (S) \\

\midrule

Bond lengths & 0.25 & 0.999 & 0.999 & 0.999 & 0.999 \\
& 0.1 & 0.947 & 0.948 & 0.949 & 0.947 \\
& 0.05 & 0.746 & 0.760 & 0.730 & 0.757 \\

\midrule

Bond angles & 0.25 & 1.000 & 1.000 & 1.000 & 1.000 \\
& 0.1 & 0.950 & 0.954 & 0.958 & 0.960 \\
& 0.05 & 0.586 & 0.564 & 0.548 & 0.519 \\

\midrule

Internal steric clash & 0.25 & 0.953 & 0.955 & 0.925 & 0.930 \\
& 0.1 &  0.776 & 0.780 & 0.778 & 0.793 \\
& 0.05 & 0.544 & 0.522 & 0.519 & 0.514 \\

\midrule

Aromatic ring flatness & 0.25 & 1.000 & 1.000 & 1.000 & 1.000 \\
& 0.1 & 1.000 & 0.999 & 1.000 & 1.000 \\
& 0.05 & 0.987 & 0.978 & 0.982 & 0.982 \\

\midrule

Double bond flatness & 0.25 & 0.993 & 0.993 & 0.993 & 0.994 \\
& 0.1 & 0.981 & 0.982 & 0.982 & 0.983 \\
& 0.05 & 0.959 & 0.961 & 0.961 & 0.962 \\

\bottomrule

\end{tabular}
\end{sc}
\end{scriptsize}
\end{center}
\vskip -0.1in
\end{table}


\end{document}